# Implementing a Cloud Platform for Autonomous Driving

Shaoshan Liu, Jie Tang, Chao Wang, Quan Wang, and Jean-Luc Gaudiot, *Fellow, IEEE*

**ABSTRACT**

Autonomous driving clouds provide essential services to support autonomous vehicles. Today these services include but not limited to distributed simulation tests for new algorithm deployment, offline deep learning model training, and High-Definition (HD) map generation. These services require infrastructure support including distributed computing, distributed storage, as well as heterogeneous computing. In this paper, we present the details of how we implement a unified autonomous driving cloud infrastructure, and how we support these services on top of this infrastructure.

## 1. INTRODUCTION

As shown in Figure 1, autonomous driving technology is a complex integration of technologies, consisting of three major divisions of R&D: the *algorithms*, including sensing, perception, and decision; the *client system*, including the robotics operating system and hardware platform; and the *cloud platform*, including data storage, simulation, high-definition (HD) map generation, and deep learning model training [1].

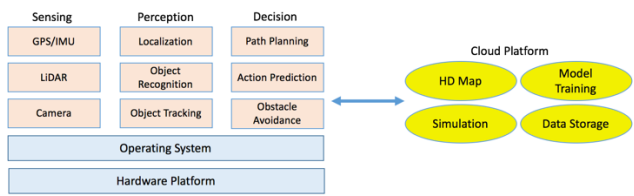

Figure 1: Architecture for Autonomous Driving

Autonomous vehicles are mobile systems, and autonomous driving clouds provide some basic infrastructure supports including distributed computing, distributed storage, and heterogeneous computing. On top of this infrastructure, we can implement essential services to support autonomous vehicles. For instance, as autonomous vehicles travel around a city, each second it can generate over 2GB of raw sensor data. It thus requires an efficient cloud infrastructure to store, and to make sense of the enormous amount of raw data. With the cloud infrastructure introduced in this paper, we can efficiently utilize the raw data to perform distributed simulation tests for new algorithm deployment, to perform offline deep learning model training, as well as to generate HD map.

## 2. INFRASTRUCTURE

The key cloud computing applications for autonomous driving include but are not limited to simulation tests for new algorithm deployment, HD map generation, and offline deep learning model training. These applications all require infrastructural support, such as distributed computing and storage. One way to do this is to tailor an infrastructure to each application, at the cost of several practical problems:

- *Lack of dynamic resource sharing*: if we tailored each infrastructure to one application, then we could not use them interchangeably even when one is idle and the other is fully loaded.
- *Performance degradation*: data is sometimes shared across applications. For instance, a newly generated map can be used in the driving simulation workloads. Without a unified infrastructure, we often need to copy data from one distributed storage element to another, leading to high performance overhead.
- *Management overheads*: it may take a team of engineers to maintain each specialized infrastructure. By unifying the infrastructure, we would greatly reduce the management overhead.

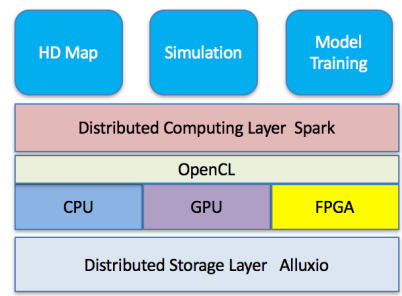

Figure 2: Cloud Platform for Autonomous Driving

As shown in Figure 2, to address these problems, we developed a unified infrastructure to provide distributed computing and distributed storage capabilities. To further improve performance, we built a heterogeneous computing layer to accelerate different kernels on GPUs or FPGAs, which either provide better performance or energy efficiency. We use Spark for distributed computing [2], OpenCL for heterogeneous computing acceleration [4], and Alluxio for in-memory storage [3]. By combining the advantages of these three infrastructure components, we can deliver a reliable, low-latency, and high-throughput autonomous driving cloud.

## 2.1 Distributed Computing Framework

When we started building the distributed computing framework for autonomous driving, we had two options, the Hadoop MapReduce engine [11], which has a proven track record, or Apache Spark [2], an in-memory distributed computing framework that provides low latency and high throughput.

Specifically, Apache Spark provides programmers with an application programming interface centered on a data structure called the resilient distributed dataset (RDD), a read-only multiset of data items distributed over a cluster of machines maintained in a fault-tolerant way. It was developed in response to limitations in the MapReduce cluster computing paradigm, which forces a particular linear dataflow structure on distributed programs: MapReduce programs read input data from disk, map a function across the data, reduce the results of the map, and store reduction results on disk. In contrast, Spark's RDDs function as a working set for distributed programs that offer a restricted form of distributed shared memory. By using in-memory RDD, Spark can reduce the latency of iterative computation by several orders of magnitude.

Before switching to Spark from MapReduce, we focused on the reliability of the Spark cluster to determine whether it can deliver the needed performance improvement. First, to verify its reliability, we deployed a 1,000-machine Spark cluster and stress-tested it for three months. The stress test helped us identify a few bugs in the system, mostly in system memory management that caused the Spark nodes to crash. After fixing these bugs, the system ran smoothly for several weeks with very few crashes, this confirmed our belief that Spark could be a viable solution for distributed computing platform for autonomous driving.

Second, to quantify performance, we ran a high number of production SQL queries on MapReduce and on a Spark cluster. With the same amount of computing resources, Spark outperformed MapReduce by 5X on average. Using an internal query that we performed daily at Baidu, it took MapReduce more than 1,000 seconds to complete, but it only took Spark 150 seconds to complete.

## 2.2 Distributed Storage

After selecting a distributed computing engine, we needed to decide on the distributed storage engine. Again, we faced two options, to remain with the Hadoop Distributed File System (HDFS) [11], which provides reliable persistent storage, or to use Alluxio, a memory-centric distributed storage system, enabling reliable data sharing at memory-speed, across cluster frameworks [3].

Specifically, Alluxio utilizes memory as the default storage medium and delivers memory-speed read and write performance. However, memory is a scarce resource and thus Alluxio may not provide enough storage space to store all the data.

The space requirement can be fulfilled by Alluxio's tiered storage feature. With tiered storage, Alluxio can manage multiple storage layers including Memory, SSD, and HDD. Using tiered storage, Alluxio can store more data in the system at the same time, since memory capacity may be limited in some deployments. Alluxio automatically manages blocks between all the configured tiers, so users and administrators do not have to manually manage the locations of the data. In a way, the Memory layer of the tiered storage serves as the top level cache, SSD serves as the second level cache, HDD serves as the third level cache, while persistent storage is the last level storage.

In our environment, we co-locate Alluxio with the compute nodes, and have Alluxio as a cache layer to exploit spatial locality. As a result, the compute nodes can read from and write to Alluxio; Alluxio then asynchronously persists data into the remote storage nodes. Using this technique, we managed to achieve a 30X speed up when compared to using HDFS only.

## 2.3 Heterogeneous Computing

By default, the Spark distributed computing framework uses a generic CPU as its computing substrate, which, however, may not be the best for certain type of workloads. For instance, GPUs inherently provide enormous data parallelism, highly suitable for high-density computations, such as convolutions on images. For instance, we have compared the performance of GPU vs. CPU on Convolution Neural Network-based object recognition tasks, and found that GPU can easily outperform CPU by a factor of 10 ~ 20 X. On the other hand, FPGA is a low-power solution for vector computation, which is usually the core of computer vision and deep learning tasks. Utilizing these heterogeneous computing substrates will greatly improve performance as well as energy efficiency.

There are several challenges on integrating these heterogeneous computing resources into our infrastructure: first, how to dynamically allocate different computing resources for different workloads. Second, how to seamlessly dispatch a workload to a computing substrate.

As shown in Figure 3, to address the first problem, we used YARN and Linux Container (LXC) for job scheduling and dispatch. YARN provides resource management and scheduling capabilities for distributed computing systems, allowing multiple jobs to share a cluster efficiently. LXC is an operating-system-level virtualization method for running multiple isolated Linux systems on the same host. LXC allows isolation, limitation, and prioritization of resources, including CPU, memory, block I/O, network, *etc*. Using LXC, one can effectively co-locate multiple virtual machines on the same host with very low overhead. Our experiments show that the CPU overhead of hosting a LXC is less than 5% comparing to running an application natively.

When a Spark application is launched, it can request heterogeneous computing resources through YARN. YARN then allocates LXCs to satisfy the request. Note that each Spark worker can host multiple containers, and that each may contain CPU, GPU, or FPGA computing resources. In this case, containers provide resource isolation to facilitate high resource utilization as well as task management.

To solve the second problem, we needed a mechanism to seamlessly connect the Spark infrastructure with these heterogeneous computing resources. Since Spark uses Java Virtual Machine (JVM) by default, the first challenge is to deploy workloads to the native space. As mentioned before, since the Spark programming interface centered on RDD, we developed a heterogeneous computing RDD which could dispatch computing tasks from the managed space to the native space through the Java Native Interface (JNI).

Next, in the native environment, we needed a mechanism to dispatch workloads to GPU or FPGA, for which we chose to use OpenCL due to its availability on different heterogeneous computing platforms. Functions executed on an OpenCL device are called kernels. OpenCL defines an API that allows programs running on the host to launch kernels on the heterogeneous devices and manage device memory.

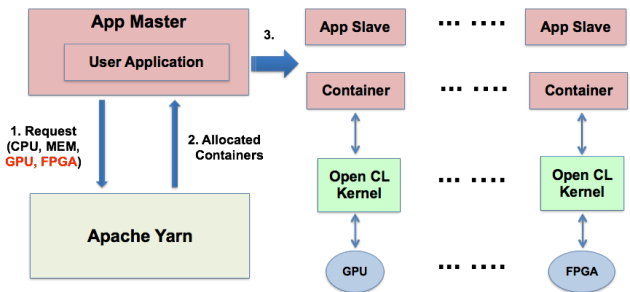

Figure 3: Distributed Heterogeneous Computing Platform

## 3. SIMULATION

With the unified infrastructure ready, let us now examine the services running on top of it. The first service we examine is distributed simulation tests for new algorithm deployment.

Whenever we develop a new algorithm, we need to test it thoroughly before we can deploy it on real cars, lest the testing cost is enormous and the turn-around time too high. Therefore, we can test the system on simulators [5]. One simulation approach consists in replaying the data through Robot Operating System (ROS) [6], where the newly developed algorithms are deployed for quick verification and early problem identification. Only after an algorithm passes all simulation tests can it be qualified to deploy on an actual car for on-road testing.

If we were to test the new algorithm on a single machine, it would either take too long or we would not have enough test coverage. To solve this problem, we leverage the Spark infrastructure to build a distributed simulation platform. This allows us to deploy the new algorithm on many compute nodes, feed each node with different chunks of data, and, at the end, aggregate the test results.

To seamlessly connect ROS and Spark, we needed to solve two problems: first, Spark by default consumes structured text data. However, for simulation we need Spark to consume multimedia binary data recorded by ROS such as raw or filtered readings from various sensors, detected obstacle bounding boxes from perception. Second, ROS needs to be launched in the native environment, where Spark lives in the managed environment.

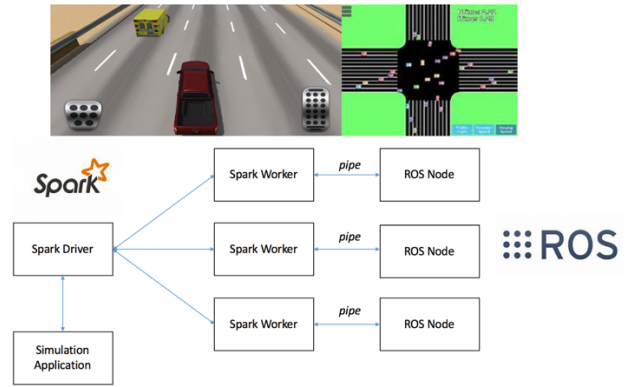

Figure 4: Simulation Platform for Autonomous Driving

### 3.1 BinPipeRDD

To make this architecture work, the first task is to have Spark consume binary input stream such as multimedia data. In the original design of Spark, inputs are in text format. Under such a context, we can have input records, as an example, with keys and values separated by space/tab characters, and records separated by Carriage Return characters. However, such an assumption is no longer valid in the context of binary data streams in which each data element in a key/value field could be of any value. To tackle this problem, we designed and implemented BinPipeRDD. Figure 5 shows how BinPipeRDD works in a Spark executor. First, the partitions of binary files go through encoding and serialization stages to form a binary byte stream. The encoding stage will encode all supported inputs format including strings (*e.g.*, file name) and integers (*e.g.*, binary content size) into our uniform format, which is based on byte array. Afterward, the serialization stage will combine all bytes arrays (each may correspond to one input binary file) into one single binary stream. Then, the user program, upon receiving that binary stream, would de-serialize and decode it according to interpret the byte stream into an understandable format. Next, the user program would perform the target computation (User Logic), which ranges from simple tasks such as rotate the jpg file by 90 degrees if needed, to relatively complex tasks such as detecting pedestrians given the binary sensor readings from LiDAR scanners. The output would then be encoded and serialized before being passed in the form of RDD[Bytes] partitions. In the last stage, the partitions can be returned to the Spark driver through a collect operation or be stored in HDFS as binary files. With this process, we can now process and transform binary data into a user-defined format and transform the output of the Spark computation into a byte stream for collect operations or take it one step further to convert the byte stream into text or generic binary files in HDFS according to the needs and logic of applications.

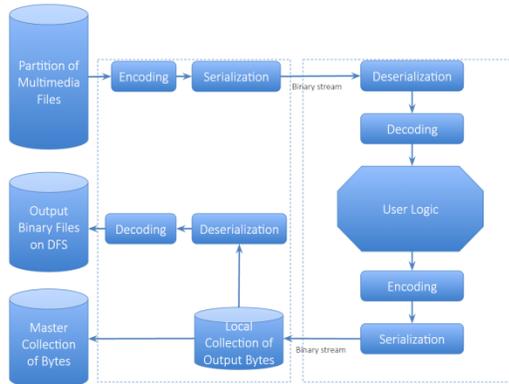

Figure 5: BinPipeRDD Design

## 3.2 Connecting Spark and ROS

With BinPipeRDD, Spark can now consume ROS Bag data, and we needed a way to launch ROS nodes in Spark as well as a way to communicate between Spark and ROS nodes. One choice was to design a new form of RDD to integrate ROS nodes and Spark, but this might involve changing ROS's as well as Spark's interfaces. Worrying about maintaining different versions of ROS, we went for a different solution and launched ROS and Spark independently, while co-locating the ROS nodes and Spark executors, and having Spark communicate with ROS nodes through Linux pipes. Linux pipes create a unidirectional data channel that can be used for inter-process communication. Data written to the write end of the pipe is buffered by the kernel until it is read from the read end of the pipe.

## 3.3 Performance

As we developed the system, we continually evaluated its performance. First, we performed basic image feature extraction tasks on one million images (total dataset size > 12 TB) and tested the system's scalability. As shown in Figure 6, as we scaled from 2,000 CPU cores to 10,000, the execution time dropped from 130 seconds to about 32 seconds, demonstrating extremely promising capability of linear scalability. Next we ran an internal replay simulation test set. On a single node, it takes about 3 hours to finish the whole dataset. As we scale to eight Spark nodes, it only takes about 25 minutes to finish the simulation, again demonstrating excellent potential for scalability.

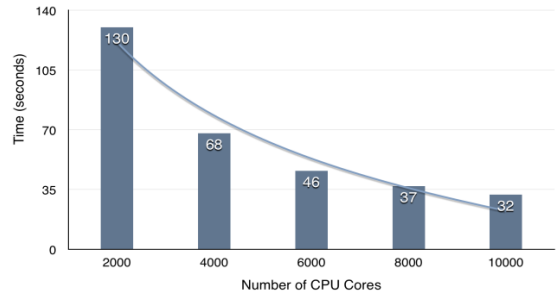

Figure 6: Simulation Platform Data Pipeline

## 4. MODEL TRAINING

The second application this infrastructure needs to support is offline model training. To achieve high performance in offline model training, our infrastructure provides seamless GPU acceleration as well as in-memory storage support of parameter servers.

As we use different deep learning models in autonomous driving, it is imperative to provide updates that will continuously improve the effectiveness and efficiency of these models. However, since the amount of raw data generated is enormous, we would not be able to achieve fast model training using single servers. To approach this problem, we developed a highly scalable distributed deep learning system using Spark and Paddle [10]. In the Spark driver, we can manage a Spark context and a Paddle context, and in each node, the Spark executor hosts a Paddler trainer instance. On top of that, we can use Alluxio as a parameter server for this system. Using this system, we have achieved linear performance scaling, even as we add more resources, proving that the system is highly scalable.

### 4.1 Why Use Spark?

The first question one may ask is why use Spark as the distributed computing framework for offline training, given that the existing deep learning frameworks all have distributed training capabilities. The main reason is that although model training looks like a standalone process, it may depend on the data preprocessing stage, such as ETL and simple feature extraction *etc*. As shown on the left side of Figure 7 below, in our practical tests, if we treated each stage as standalone, this would involve intensive I/O to the underlying storage, such as HDFS. As a consequence, we discovered that the I/O to the underlying storage often became the bottleneck of our whole processing pipeline.

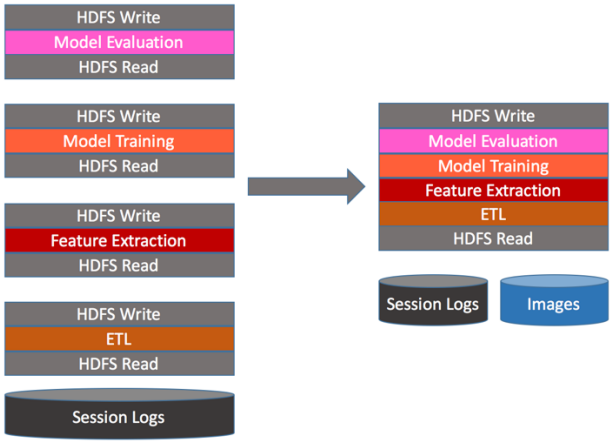

Figure 7: Training Platform for Autonomous Driving

As shown on the right side of Figure 7, by using Spark as the unified distributed computing framework, we can now buffer the intermediate data in memory, in the form of RDDs. The processing stages naturally form a pipeline without intensive remote IO accesses to the underlying storage in between the stages. This way, we read the raw data from HDFS at the beginning of the pipeline, and then pass the processed data to the next stage in the form of RDDs, until we finish the last stage and at last write the data back to HDFS. This approach allowed us to effectively double, on average, the throughput of the system.

## 4.2 Training Platform Architecture

Figure 8 shows the architecture of our training platform. First, we have a Spark driver to manage all the Spark nodes, with each node hosts a Spark executor and a Paddle trainer, which allows us to utilize the Spark framework to handle distributed computing and resource allocation.

With this architecture, we can exploit data parallelism by partitioning all training data into shards so that each node independently processes one or more shards of the raw data. To synchronize the nodes, at the end of each training iteration, we need to summarize all the parameter updates from each node, perform calculations to derive a new set of parameters, and then broadcast the new set of parameters to each node so they can start the next iteration of training.

It is the role of the parameter server to efficiently store and update the parameters. If we were to store the parameters in HDFS, then again, as we have alluded to earlier, I/O would become the performance bottleneck. To alleviate this problem, we utilized Alluxio as our parameter server. As shown in section 2.2, Alluxio is a memory-centric distributed storage, which utilizes in-memory storage to optimize for its I/O performance. Comparing to HDFS, we have observed an I/O performance gain factor of more than 5X by utilizing Alluxio as parameter servers.

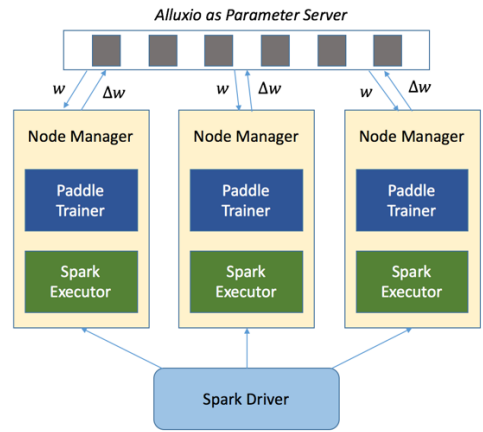

Figure 8: Training Platform for Autonomous Driving

## 4.3 Heterogeneous Computing

Next, we explored how heterogeneous computing could improve the efficiency of offline model training. As a first step, we explored how GPU performed compared to a CPU with Convolution Neural Networks (CNN). Using an internal object recognition model with the OpenCL infrastructure presented in section 2.3, we have observed a 15X speed-up using GPU. The second step was to understand the scalability of this infrastructure. On our machine, each node is equipped with one GPU card. Figure 9 shows the result of this study, as we scaled the number of GPUs, the training latency per pass dropped almost linearly. This result confirmed the scalability of our platform, such that as we have more data to train against, we could reduce the training time by providing it with more computing resources.

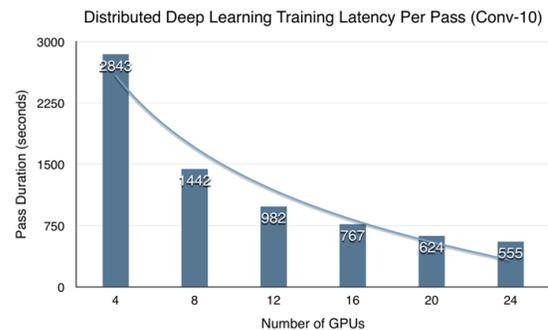

Figure 9: Performance of Distributed Model Training

## 5. HD MAP GENERATION

The third application this infrastructure needs to support is HD map generation, a multi-stage pipeline. By using Spark and heterogeneous computing, we managed to reduce the IO between the pipeline stages and accelerate the critical path of the pipeline.

As shown in Figure 10, like offline training, HD map production is also a complex process that involves many stages, including raw data reading, filtering and preprocessing, pose recovery and refinement, point cloud alignment, 2D reflectance map generation, HD map labeling, as well as the final map outputs [7, 8]. Using Spark, we can connect all these stages together in one Spark job. A great advantage is that Spark provides an in-memory computing mechanism, such that we do not have to store the intermediate data in hard disk, thus greatly reducing the performance of the map production process.

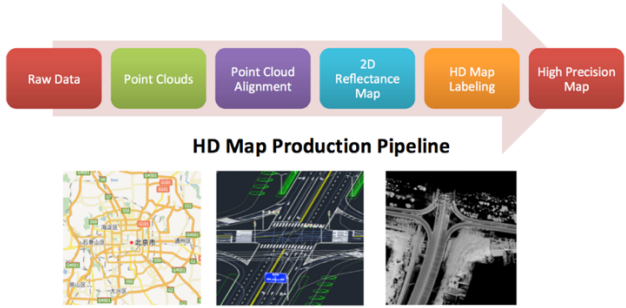

Figure 10: Simulation Platform for Autonomous Driving

## 5.1 HD Map

Just as with traditional digital maps, HD maps have many layers of information. As shown in Figure 11, at the bottom layer we have a grid map generated by raw LiDAR data, with a grid granularity of about 5 centimeters by 5. This grid basically records elevation and reflection information of the environment in each grid cell. As the autonomous vehicles are moving and collecting new LiDAR scans, they compare in real time the new LiDAR scans against the grid map with initial position estimates provided by GPS and/or IMU, which then assists these vehicles in precisely self-localizing in real-time.

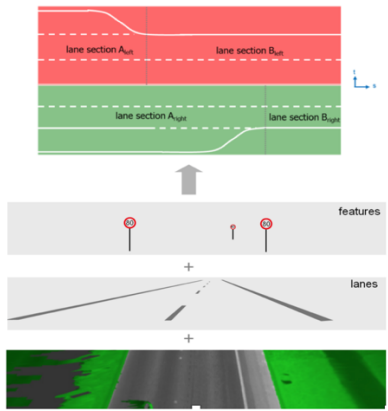

Figure 11: Performance of Distributed Model Training

On top of the grid layer, there are several layers of semantic information. For instance, the reference line and lane information are added to the grid map to label each lane. This allows autonomous vehicles to determine whether they are on the correct lane when moving, and to also decide whether they are maintaining a safe distance to the vehicles on neighboring lanes. On top of the lane information, traffic sign labels will be added to notify the autonomous vehicles of the current speed limit, and whether traffic lights are nearby *etc*. This gives an additional layer of protection in case the sensors on the autonomous vehicles fail to catch the signs.

## 5.2 Map Generation in the Cloud

Although we mentioned the importance of LiDAR data in HD map generation, it is not the only sensor data used. As shown in Figure 12, the HD map generation process actually fuses raw data from multiple sensors in order to derive accurate position information. First, the wheel odometry data and the IMU data can be used to perform propagation, or to derive the displacement of the vehicle within a fixed amount of time. Then the GPS data and the LiDAR data can be used to correct the propagation results in order to minimize errors.

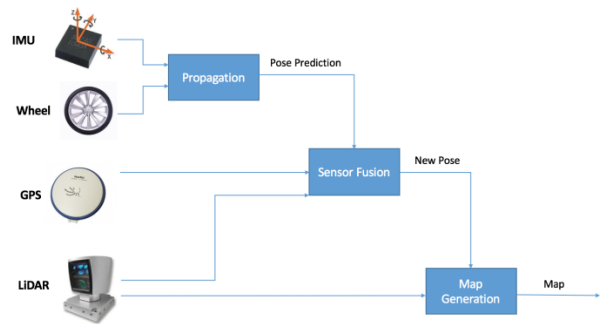

Figure 12: Map Generation in the Cloud

In terms of process, the computation of map generation can be divided into three stages: first, Simultaneous Localization And Mapping (SLAM) is performed to derive the location of the each LiDAR scan. In this stage, the Spark job loads all the raw data, including IMU log, wheel odometry log, GPS log, and LiDAR raw data from HDFS. Second, it performs map generation and point cloud alignment, in which the independent LiDAR scans are stitched together to form a continuous map. Third, label and semantic information is added to the grid map.

Just as with offline training applications, we linked these stages together using a Spark job and buffered the intermediate data in memory. By using this approach, we achieved a 5X speedup when compared to having separate jobs for each stage. Also, the most expensive operation for the map generation stage is the iterative closest point (ICP) point cloud alignment [12]. By using the heterogeneous infrastructure, we managed to accelerate this stage by 30X by offloading the core of ICP operations to GPU.

## 6. CONCLUSIONS

An autonomous driving cloud is an essential part of the autonomous driving technology stack. In this paper, we have shown the details of our practical experiences of building a production autonomous driving cloud. To support different cloud applications, we need an infrastructure to provide distributed computing, distributed storage, as well as hardware acceleration through heterogeneous computing capabilities.

If we were to tailor the infrastructure for each application, we would have to maintain multiple infrastructures, potentially leading to low resource utilization, low performance, and high management overhead. We solved this problem by building a unified infrastructure with Spark for distributed computing, Alluxio for distributed storage, and OpenCL to exploit heterogeneous computing resources for further performance improvement and energy efficiency.

With the unified infrastructure, many applications can be supported, including but not limited to distributed simulation tests for new algorithm deployment, offline deep learning model training, and HD map generation. We have delved into each of these applications to explain how the infrastructure can be utilized to support the specific features, and to provide performance improvement as well as scalability.

At this point, we are in the early stages of the development of a cloud infrastructure for autonomous vehicles, as autonomous driving technologies are actively evolving. Nonetheless, we know that, by having a unified infrastructure to provide the basic capabilities, including distributed computing, distributed storage, and heterogeneous computing, autonomous driving cloud itself can quickly evolve to meet the needs of emerging autonomous driving cloud applications.

## 7. ACKNOWLEDGEMENTS

This work is partly supported by the National Science Foundation under Grant No. XPS-1439165. Any opinions, findings, and conclusions or recommendations expressed in this material are those of the authors and do not necessarily reflect the views of NSF.

Dr. Shaoshan Liu is the co-founder of PerceptIn. He attended UC Irvine for his undergraduate and graduate studies and obtained a Ph.D. in Computer Engineering in 2010. His research focuses on Computer Architecture, Big Data Platforms, Deep Learning Infrastructure, and Robotics. He has over eight years of industry experience: before co-founding PerceptIn, he was with Baidu USA, where he led the Autonomous Driving Systems team. Before joining Baidu USA, he worked on Big Data platforms at LinkedIn, Operating Systems kernel at Microsoft, Reconfigurable Computing at Microsoft Research, GPU Computing at INRIA (France), Runtime Systems at Intel Research, and Hardware at Broadcom. Email: shaoshan.liu@perceptin.io

Dr. Jie Tang is the corresponding author and she is currently an associate professor in the School of Computer Science and Engineering of South China University of Technology, Guangzhou, China. Before joining SCUT, Dr. Tang was a post-doctoral researcher at the University of California, Riverside and Clarkson University from Dec. 2013 to Aug. 2015. She received the B.E. from the University of Defense Technology in 2006, and the Ph.D. degree from the Beijing Institute of Technology in 2012, both in Computer Science. From 2009 to 2011, she was a visiting researcher at the PArallel Systems and Computer Architecture Lab at the University of California, Irvine, USA. Email: cstangjie@scut.edu.cn

Chao Wang is currently a Senior Software Architect at Baidu Autonomous Driving Unit, focusing on distributed simulation platform. He received a Master Degree in Computer Science from University of Southern California. He has over five years of industry experience. His primary interest is in Cloud Computing and Big Data Platforms. Email: wangchao30@baidu.com

Dr. Quan Wang is currently a Principal Architect at Baidu Autonomous Driving Unit, focusing on high-definition map generation. He received a Ph.D. in Computer Science from University of Southern California. He has over ten years of industry experience. His primary research interests include computer vision, and distributed computing. Email: wangquan02@baidu.com

Dr. Jean-Luc Gaudiot received the Diplôme d'Ingénieur from ESIEE, Paris, France in 1976 and the M.S. and Ph.D. degrees in Computer Science from UCLA in 1977 and 1982, respectively. He is currently Professor in the Electrical Engineering and Computer Science Department at UC, Irvine. Prior to joining UCI in 2002, he was Professor of Electrical Engineering at the University of Southern California since 1982. His research interests include multithreaded architectures, fault-tolerant multiprocessors, and implementation of reconfigurable architectures. He has published over 250 journal and conference papers. His research has been sponsored by NSF, DoE, and DARPA, as well as a number of industrial companies. He has served the community in various positions and was elected to the presidency of the IEEE Computer Society for 2017. E-mail: gaudiot@uci.edu